\documentclass[epj]{svjour}

\usepackage{graphics}
\usepackage{color}
\usepackage{soul}
\usepackage{ulem}
\usepackage{graphicx}
\usepackage{dcolumn}
\usepackage{bm}
\usepackage[T1]{fontenc}
\usepackage{mathptmx}
\usepackage{amsmath}
\begin{document}
\title{Near-threshold photoproduction of $J/\psi$ in two-gluon exchange model}

\author{Fancong Zeng\inst{1},
Xiao-Yun Wang\inst{1,2}\thanks{xywang@lut.edu.cn (corresponding author)},
Li Zhang\inst{1,2},
Ya-Ping Xie\inst{3} \thanks{xieyaping@impcas.ac.cn},
Rong Wang\inst{3} \thanks{rwang@impcas.ac.cn},
and Xurong Chen\inst{3} \thanks{xchen@impcas.ac.cn}}

\institute{Lanzhou University of Technology, Lanzhou 730050, China
\and Department of physics, Lanzhou University of Technology
\and Institute of Modern Physics, Chinese Academy of Sciences, Lanzhou 730000, China}

\date{\today}
%
\abstract{The near-threshold photoproduction of $J/\psi$ is regarded as one golden process to
unveil the nucleon mass structure, pentaquark state involving the charm quarks,
and the poorly constrained gluon distribution of the nucleon at large $x$ ($>0.1$).
In this paper, we present an analysis of the current experimental data
under a two-gluon exchange model, which shows a good consistency.
Using a parameterized function form with three free parameters, we have determined the nucleonic gluon
distribution at the $J/\psi$ mass scale. Moreover, we predict the differential
cross section of the electroproduction of $J/\psi$ as a function of the invariant mass of the final
hadrons $W$, at EicC, as a practical application of the model and the obtained gluon distribution.
According to our estimation, thousands of $J/\psi$ events can be detected per year on EicC near the threshold.
Therefore, the relevant experimental measurements are suggested to be carried out on EicC.
} 
\authorrunning{Title}
\maketitle
\section{Introduction}
\label{sec:intro}

The photoproduction of $J/\psi$ close to the threshold is a key experimental channel widely discussed
to investigate the pentaquark photoproduction, for the charm production near the threshold has a strong sensitivity
to the multi-quark, the gluonic and the hidden-color correlations to the hadronic and nuclear wavefunctions in QCD \cite{Brodsky:2000zc}.
Moreover, the near-threshold photoproduction of $J/\psi$ also plays an important role
in probing the nucleon mass structure \cite{Ji:1994av,Ji:1995sv,Lorce:2017xzd,Kharzeev:1995ij,Kharzeev:1998bz},
as has been recently illustrated with a very preliminary analysis \cite{Wang:2019mza}
of the GlueX data \cite{Ali:2019lzf}.
On the experimental side, there has been some progresses reported and undergoing \cite{Ali:2019lzf,Collaboration,Meziani:2016lhg,Joosten:2018gyo}.

Recently, the first measurement of the near-threshold cross section of the reaction $\gamma p \rightarrow J / \psi p$
has been reported \cite{Collaboration}. Including the GlueX data, the photon-gluon fusion model and the pomeron exchange model \cite{Xu:2020uaa}
has been demonstrated to be applicable to explain the heavy quarkonia photoproduction in a wide energy range.
More sophisticated models based on the three-gluon exchange, the holographic QCD,
and the dispersion relation are developed \cite{Hatta:2018ina,Hatta:2019lxo,Mamo:2019mka,Boussarie:2020vmu,Gryniuk:2020mlh},
and taken to explain the recent data.
Unfortunately, it is difficult for these models to give the differential cross section at the production threshold,
where some variations of the predictions are found.
There are many models that can describe the photoproduction of $\gamma p \rightarrow J / \psi p$
successfully in different energy ranges.
Among them, the two-gluon exchange model captures our attention.
One work\cite{Hatta:2018ina}  shows the total and differential cross sections compared with the experimental data,
providing some ideas and literatures for us to do this work, confronting with the recent experimental data.
The exclusive photoproductions of all vector mesons by real and virtual photos
are studied in a soft dipole Pomeron model \cite{Martynov:2002ez},
which has a perfect quality of fitting to both the total and the differential cross sections
in the high energy region. However, there is an inconsistency for the differential cross section near the threshold.
In a related work\cite{Cao:2019kst},  the contribution of pentaquark state $P_c$ was added,
which is based on the nonresonant contribution parametrized with the soft dipole Pomeron model.

The total elastic $J / \psi - p$ production at high photon-nucleon invariant mass $W$ is well described
by the $t-$channel exchange of a colorless object between the photon and the proton \cite{Frankfurt:2002ka}.
In this paper, two-gluon exchange model is applied to fit the data by GlueX Collaboration\cite{Ali:2019lzf}.
This makes the total and differential cross sections depend on the gluon distribution function squared,
while the conventional gluon distribution function from GRV98 \cite{Gluck:1998xa}, NNPDF \cite{Ball:2011uy},
CJ15 \cite{Owens:2012bv,Accardi:2016qay}, and IMParton16 \cite{Wang:2016sfq} are difficult to interpret the new GlueX data \cite{Ali:2019lzf},
within the framework of the two-gluon exchange model.
Thus we plan to use a simplified gluon distribution parametrization \cite{Pumplin:2002vw} that has the form
$xg\left(x, m_{J}^{2}\right)=A_0 x^{A_1}(1-x)^{A_2}$ to perform a fit to the near-threshold $J/\psi$ photoproduction data.
The other purpose of this work is to predict the total and differential cross sections of $J/\psi$ electroproduction
at different energies near the threshold, based on the obtained gluon distribution,
in order to remove some model uncertainties from the gluon distribution.
Nowadays, the high and low energy Electron Ion Colliders (EIC)
are vigorously proposed to be built for probing the deepest structure inside hadron,
which is the main building block of the visible universe.
US EIC is on the way and focus on the high energy collisions \cite{Accardi:2012qut},
and the opportunities of Chinese EIC are now under some hot discussions \cite{Chen:2018wyz,Chen:2019equ}.
To make predictions for the future machines are necessary.

The paper is organized as follows.
The formulas of the two-gluon exchange model and the $ J / \psi$ production via electron-proton collisions
are provided in Sec. \ref{sec:formalism}.
Then in Sec. \ref{sec:results}, we show the numerical result on the explanations of the current experimental data,
the extracted gluon distribution, and the predictions of $J/\psi$ production on EicC \cite{Chen:2018wyz,Chen:2019equ}.
A short summary is given in Sec. \ref{sec:summary}.

\section{Formalism}
\label{sec:formalism}

\begin{figure}[h!]
\center
\begin{minipage}[t]{0.45\textwidth}
\includegraphics[scale=0.6]{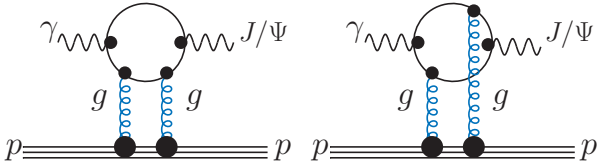}
\end{minipage}
\caption{The schematic Feynman diagram of the two-gluon exchange model for $J/\psi$ production.}
\label{fig:TGE-model}
\end{figure}

The two-gluon exchange model is based on the photon fluctuation into the quark-antiquark pair ($\gamma \rightarrow q+\bar{q}$)
and the picture of the double gluon exchange between the nucleon state and the quark-antiquark pair,
which is illustrated in Fig. \ref{fig:TGE-model}.
Due to the hard scale in the heavy quarkonium production, the exclusive vector meson photoproduction amplitude
is factorized as a reasonable assumption.
In a lowest order perturbative QCD of factorization, the photoproduction of $J/\psi$ amplitude is given by
\cite{Ryskin:1992ui,Brodsky:1994kf,Ryskin:1995hz},

\begin{equation}
\label{eq:amplitudeT}
\mathcal{T}=\frac{i 2 \sqrt{2} \pi^{2}}{3} m_{q} \alpha_{s} e_{q} f_{V} F_{2 g}(t) \int d l^{2} D_{g}^{2}(l)\left[D_{+}(l)-D_{-}(l)\right] G(l).
\end{equation}
The mason decay constant $f_V$ is deduced from the leptonic decay width, which is given by,

\begin{equation}
\Gamma_{e^{+} e^{-}}=\frac{8 \pi \alpha^{2} e_{q}^{2}}{3 m_{V}} f_V^{2}.
\end{equation}
The gluon propagator $D_g(l)$ in Eq. (\ref{eq:amplitudeT}) is taken to be $1/l^2$.
$D_-(l)$ represents the propagator of the off-shell quark when the two gluons
couple to different quarks of the vector meson, which is written as,

\begin{equation}
D_{-}(l)=\left(-2 m_{q}^{2}-2 l^{2}\right)^{-1},
\end{equation}
while $D_+(l)$ represents the propagator of the off-shell quark when the two gluons
couple to the same quark in the vector meson, which is written as,

\begin{equation}
D_{+}(l)=\left(-2 m_{q}^{2}\right)^{-1}.
\end{equation}
The factor $F_{2g}(t)$ in Eq. (\ref{eq:amplitudeT}) accounts for the dependence of the amplitude
with respect to the two gluon correlation in the proton, and it is written as \cite{Ryskin:1992ui,Sibirtsev:2004ca},

\begin{equation}
F_{2g}(t)=\frac{4 m_{p}^{2}-2.8 t}{4 m_{p}^{2}-t} \frac{1}{\left(1-t / t_{0}\right)^{2}},
\end{equation}
in which $t_0=0.71$ GeV$^2$.
$G(l)$ defines the probability for the dipole to catch the gluon
of momentum $l$ from the proton.
Its integral is related to the gluon distribution function $xg(x)$,
which is written as \cite{Ryskin:1992ui,Brodsky:1994kf},
\begin{equation}
x g\left(x, Q^{2}\right)=\int d l^{2} \frac{G(l)}{l^{2}}.
\end{equation}

With the above discussions, the two-gluon exchange amplitude becomes \cite{Sibirtsev:2004ca},

\begin{equation}
\begin{split}
\mathcal{T}=&\frac{i \sqrt{2} \pi^{2}}{3} m_{q} \alpha_{s} e_{q} f_{V} F_{2 g}(t)[\frac{x g\left(x, Q_{0}^{2}\right)}{m_{q}^{4}}   \\
&+\int_{Q_{0}^{2}}^{+\infty} \frac{d l^{2}}{m_{q}^{2}\left(m_{q}^{2}+l^{2}\right)} \frac{\partial x g\left(x, l^{2}\right)}{\partial l^{2}}].
\end{split}
\end{equation}
The amplitude is normalized and $\frac{d \sigma}{d t}=\alpha |\mathcal{T}|^{2}$.
In the lowest order, the $J/\psi$ photoproduction cross section is given as\cite{Sibirtsev:2004ca},

\begin{equation}
\label{eq:diff-crosssection}
\frac{d \sigma}{d t}=\frac{\pi^{3} \Gamma_{e^{+}e^{-}} \alpha_{s}}{6 \alpha m_{q}^{5}}\left[x g\left(x, m_{J}^{2}\right)\right]^{2} \exp (- b t),
\end{equation}
where $x=m_J^2/W^2$, $\alpha=1/137$ is the electromagnetic coupling constant,
$\alpha_{s}$ is the QCD coupling constant\cite{Xu:2020uaa},
and the mass of the charm quark is $m_q=1.27$ GeV.
 The radiative decay width $\Gamma_{e^+e^-}=5.547$ keV is taken from PDG average.
The exponential slope $b$ of $t$-dependence is found to be $1.67 \pm 0.38\mathrm{GeV}^{-2}$
at $W=4.59$ GeV \cite{Ali:2019lzf}.
Here, the $W$-dependence of the slope $b$ can be evaluated with
an empirical formula as $b=-\frac{\mathrm{d}}{\mathrm{d} t} \ln \left[\frac{\mathrm{d} \sigma}{\mathrm{d} t}\right]$.

$xg\left(x, m_{J}^{2}\right)$ is the gluon distribution function at $Q^2=m_{J}^{2}$,
and in this work it is parameterized using a simple function form $xg\left(x, m_{J}^{2}\right)=A_0 x^{A_1}(1-x)^{A_2}$\cite{Pumplin:2002vw}.
Fixing the gluon parametrization is one of our purposes in this work.
The poles at $x=0$ and $x=1$ in the parametrization are the singularities associated with Regge behavior at small $x$
and the quark counting rules at large $x$.
The parameters $A_0, A_1, A_2$ can be fixed by the experimental data.

\begin{figure}[h!]
\begin{minipage}[t]{0.01\textwidth}
\includegraphics[scale=0.4]{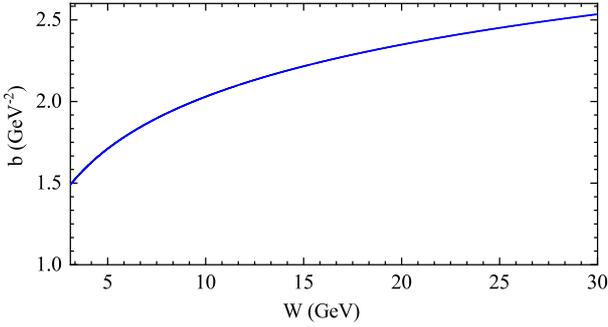}
\end{minipage}
\caption{The exponential slope $b$ as a function of the energy $W$.}
\label{fig:bslope-vs-W}
\end{figure}

The total cross section is obtained by integrating the differential cross section (Eq. (\ref{eq:diff-crosssection}))
over the allowed kinematical range from $\mid t_0\mid$ to $\mid t_1\mid$, here, $t_{0}\left(t_{1}\right)=\left[\frac{m_{1}^{2}-m_{3}^{2}-m_{2}^{2}+m_{4}^{2}}{2 W}\right]^{2}-\left(p_{1 \mathrm{cm}} \mp p_{3 \mathrm{cm}}\right)^{2}$,
$p_{i \mathrm{cm}}=\sqrt{E_{i \mathrm{cm}}^{2}-m_{i}^{2}} (i=1,3)$,
$E_{1 \mathrm{cm}}=\frac{W^2+m_{1}^{2}-m_{2}^{2}}{2 W}$,
 $E_{3 \mathrm{cm}}=\frac{W^2+m_{3}^{2}-m_{4}^{2}}{2 W}$ \cite{ww} and $W=4.59$ GeV.
 Therefore, the total cross section  can be written as,

\begin{equation}
\sigma=\frac{0.487}{b(W)} \frac{\pi^{3} \Gamma_{e^{+}e^{-}} \alpha_{s}}{6 \alpha m_{q}^{5}}  \left[x g\left(x, m_{J}^{2}\right)\right]^{2}.
\end{equation}
The exponential slope $b(W)$ has  little vary with the energy $W$ \cite{Meziani:2016lhg,Chekanov:2002xi,Adloff:2000nx,Levy:1997bh}.
Studies of S. Chekanov imply that the $b$ function in the high energy region
could be formulated as,

\begin{equation}
b(W)=b_0+0.46\cdot In(W/W_0).
\end{equation}
In order to estimate $ b(W)$  in the low energy region near threshold,
we fixed the slope to be the measured value $1.67\pm 0.38$ GeV$^{-2}$ at the enery $W_0 \simeq  4.59^{-0.15}_{+0.22}$ GeV \cite{Ali:2019lzf}.
In Fig. \ref{fig:bslope-vs-W}, it is  obvious to see that $b$ has a weak $W$-dependence
in a wide energy range. Therefore the uncertainty of $b$ is not large,
and it does not affect much the uncertainty of the total cross section.

The exclusive electroproduction of $J/\psi$ is closely connected $J/\psi$ photoproduction,
as in the electron scattering process the $J/\psi$ vector meson is generated from the virtual
photon exchanged between the electron and the hadron.
The electroproduction cross section of $J/\psi$ in electron-proton scattering
can be found in the recent literatures \cite{Lomnitz:2018juf,Klein:2019avl},

\begin{equation}
\sigma(ep\to eJ/\psi p)=\int dkdQ^2\frac{dN^2(k,Q^2)}{dkdQ^2}\sigma_{\gamma^* p\to J/\psi p}(W,Q^2),
\end{equation}
here $W$ is the center-of-mass energy of the photon-proton system,
$k$ is the momentum of the virtual photon emitted from the electron beam
in the target rest frame, and $Q^2$ is the virtuality of the photon.
The photon flux is given as \cite{Budnev:1974de},

\begin{equation} \label{eq:photoflux}
\frac{d^2N(k,Q^2)}{dkdQ^2}=\frac{\alpha}{\pi kQ^2}\Big[1-\frac{k}{E_e}+\frac{k^2}{2E^2_e}-\Big(1-\frac{k}{E_e}\Big)\Big|\frac{Q^2_{min}}{Q^2}\Big|\Big],
\end{equation}
where $E_e$ is the energy of the initial electron in proton rest frame,
and $Q^2_{min}$ is given as,

\begin{eqnarray}
Q^2_{min}=\frac{m_e^2k^2}{E_e(E_e-k)}.
\end{eqnarray}
The maximum $Q^2$ is determined by the energy loss of the initial electron,

\begin{eqnarray}
Q^2_{max}=4E_e(E_e-k).
\end{eqnarray}
The connection between the cross section induced by a real photon
and that induced by a virtual photon is governed by,
\begin{equation} \label{eq:Q2xsection}
\sigma_{\gamma^*p\to J/\psi p}(W,Q^2)=\sigma_{\gamma p\to J/\psi p}(W,Q^2=0)\bigg(\frac{M_V^2}{M_V^2+Q^2}\bigg)^\eta.
\end{equation}
in which $\eta=c_1+c_2(M_V^2+Q^2)$ with the values of $c_1 = 2.36\pm0.20$ and $c_2 =0.0029\pm 0.43\!\!\!\!\!\quad\mathrm{ GeV}^2$ \cite{Lomnitz:2018juf}.
With the known photon flux from the electron beam and the photon virtuality dependence of the cross section,
we can calculate the total cross-section in electron-proton scattering near the production threshold.

\section{Numerical results}
\label{sec:results}

\begin{figure}[h!]
\flushleft
\begin{minipage}[t]{0.01\textwidth}
\includegraphics[scale=0.4]{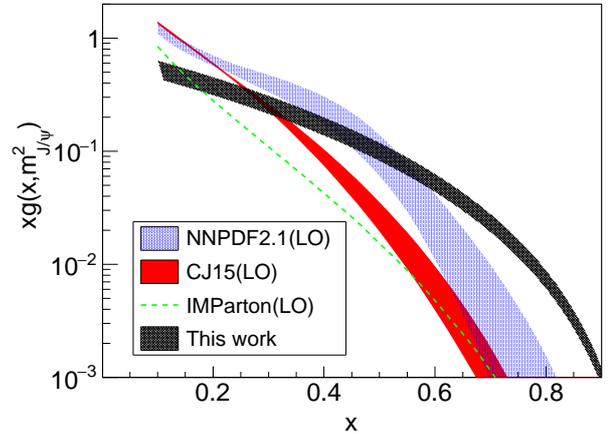}
\end{minipage}
\caption{The gluon distribution extracted from the near-threshold $J/\psi$ photoproduction data under
the two-gluon exchange model. }
\label{fig:gluonPDF}
\end{figure}

The parametrization $xg(x,m_{J}^2)=A_0 x^{A_1} (1-x)^{A_2}$ of the nucleon gluon distribution
is introduced and used in the two-gluon exchange model discussed in the above section. The free parameters $A_0, A_1, A_2$
then are fixed by a global analysis of both the total cross section data\cite{Ali:2019lzf,Chekanov:2002xi,Binkley:1981kv,Frabetti:1993ux}
and the differential cross section data near threshold ($W=4.59$ GeV) by GlueX collaboration \cite{Ali:2019lzf}.
The obtained parameters
are listed in Table \ref{tab:GluonParameters}.
Fig. \ref{fig:gluonPDF} shows our obtained gluon distribution in this work, compared with the predictions
from the widely used global fits, such as NNPDF \cite{Ball:2011uy}, CJ15 \cite{Owens:2012bv,Accardi:2016qay} and IMParton \cite{Wang:2016sfq}.
It is found that the gluon distributions determined by different groups are more or less consistent with each
other in the low $x$ range of $x<0.3$. On the other side, our obtained gluon distribution
is higher than other predictions when $x>0.6$.

\begin{table}
\footnotesize
\centering
\label{tab:1}
\caption{The fitted values of the parameters $A_0, A_1, A_2$ describing the gluon distribution function $xg(x)$
and the reduced $\chi^{2}$/d.o.f. in the low $W$ region. }     
\begin{tabular}{ccccc}
\hline\hline\noalign{\smallskip}
  $A_0$ & $A_1$ & $ A_2$ &   $\chi^{2}$/d.o.f.    \\
\noalign{\smallskip}\hline\noalign{\smallskip}
  $0.71\pm 0.12$ & $-0.00061\pm 0.00045$ & $2.83\pm0.26$ &   0.20 \\
\hline
\end{tabular}
\label{tab:GluonParameters}
\end{table}

\begin{figure}[h!]
\begin{minipage}[t]{0.1\textwidth}
\includegraphics[scale=0.4]{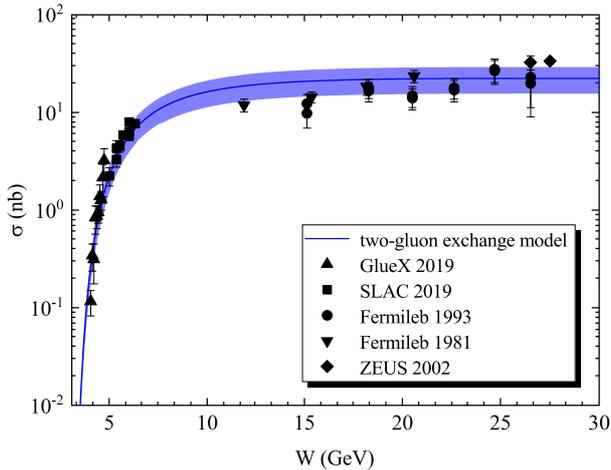}
\end{minipage}
\caption{The total cross section of the channel $\gamma p \rightarrow J / \psi p$ as a function of $W$.  }
\label{fig:JPsiTotalCrosssection}
\end{figure}

\begin{figure}[h!]
\begin{minipage}[t]{0.1\textwidth}
\includegraphics[scale=0.4]{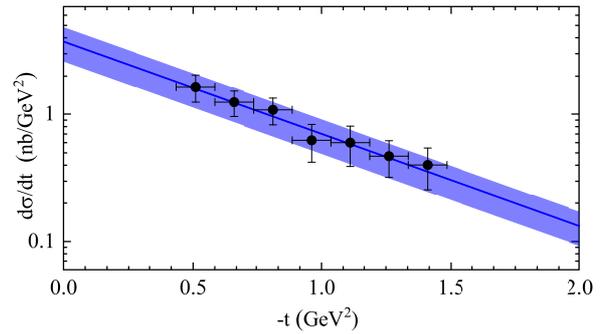}
\end{minipage}
\caption{The fitted differential cross section as a function of four momentum transfer squared $t$ at $W=4.59$ GeV
compared to the measured data by GlueX \cite{Ali:2019lzf}. }
\label{fig:JPsiDiffCrosssection}
\end{figure}

The predicted total cross section of $\gamma p \rightarrow J / \psi p$ as a function of center-of-mass
energy $W$ is shown in Fig. \ref{fig:JPsiTotalCrosssection}, compared to the experimental data from several experiments.
The comparison between the differential cross section in the two-gluon exchange model with the assumed exponential slope
and the experimental measurement of the differential cross section is manifested in Fig.\ref{fig:JPsiDiffCrosssection},
exhibiting an amazing agreement. The two-gluon exchange model is valid to describe
the $J/\psi$ photoproduction near the production threshold.

\begin{figure}[h!]
\flushleft
\begin{minipage}[t]{0.4\textwidth}
\includegraphics[scale=0.44]{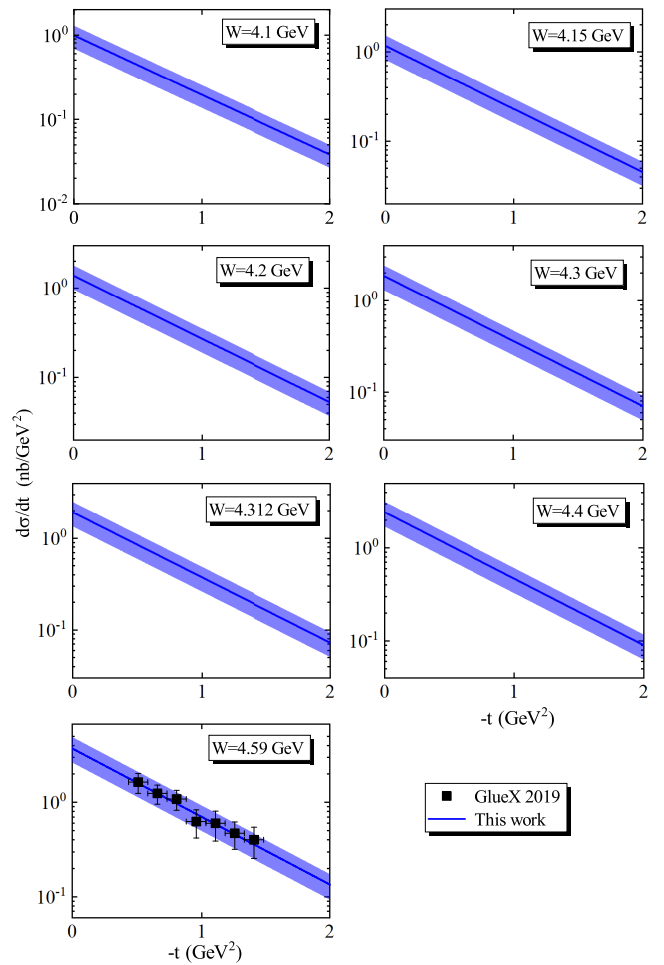}
\end{minipage}
\caption{The predicted differential cross section as a function of four momentum transfer squared $t$ at different $W$ values.  }
\label{fig:JPsiDiffCrosssectionPredictions}
\end{figure}

Since the main purpose of this paper is to search the underlying mechanism
of $J/\psi$ photoproduction near the production threshold,
the $\chi^2/N$ of the global fit in the low energy range ($W<6.4$ GeV)
is calculated to be $0.20$. The $\chi^2/N$ value less than one indicates that the two-gluon exchange model
is applicable for the near-threshold photoproduction of $J/\psi$.
Note that the errors of the experimental data from Fermilab 1981\cite{Binkley:1981kv}, Fermilab 1993\cite{Frabetti:1993ux}
and ZEUS 2002\cite{Chekanov:2002xi} are kind of big.
Hence, more future precise measurements on the near-threshold $J/\psi$ production are needed.
The $J/\psi$ photoproductions near the threshold at more different energies are predicted
by the two-gluon exchange model, which are shown in Fig. \ref{fig:JPsiDiffCrosssectionPredictions}.
The corresponding cross sections at $t=0$ GeV$^2$ are given in Table \ref{tab:ForwardCrosssections}.
These forward differential cross sections are speculated to be connected to the nucleon mass.

\begin{figure}[h!]
\flushleft
\begin{minipage}[t]{0.01\textwidth}
\includegraphics[scale=0.4]{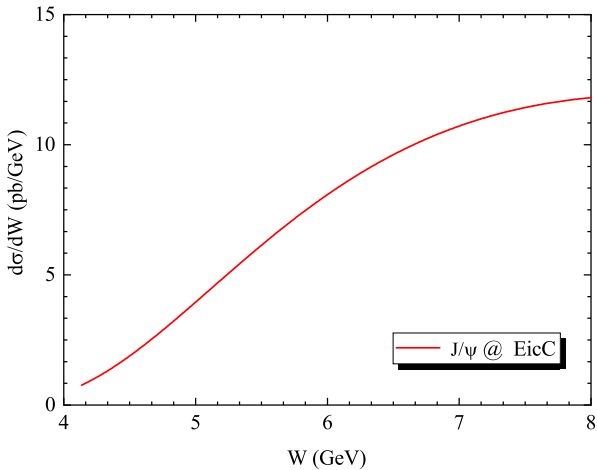}
\end{minipage}
\caption{The predictions of $J/\psi$ electro-production cross section as a function of $W$
on EicC machine, with the selections of the low virtuality photons of $0.1$ $\mathrm{ GeV}^2< Q^2 < 1$ $\mathrm{GeV}^2$.  }
\label{fig:JPsiElectroproductionAtEicC}
\end{figure}

The next generation of the advanced accelerator facility to study the $J/\psi$ photoproduction
is the electron-ion collider. The Electron ion collider in China (EicC) is under some hot discussions
and proposed to be built and run at a low energy compared to the EIC in US.
In order to investigate the opportunity of EicC in $J/\psi$ study,
we calculate the differential cross section of $J/\psi$ electroproduction as a function
of the c.m. energy of the system of the emitted virtual photon and the proton,
which is depicted in Fig. \ref{fig:JPsiElectroproductionAtEicC}.
The cross section is around a doze of pb,
which suggests a high yield rate at the high luminosity EicC.

\begin{table}
\small
\centering
\caption{The values of $d\sigma /dt|_{t=0}$ at different values of energy $W$. }      
\begin{tabular*}{\hsize}{@{}@{\extracolsep{\fill}}llllllll@{}}
\hline\hline

$W$ (GeV) & 4.1 & 4.15   & 4.2   & 4.3  &  4.312  &  4.4    &4.59   \\
\hline
 $d\sigma /dt|_{t=0}$   & 0.98 & 1.17   & 1.38  & 1.86 & 1.93  & 2.44 &3.73 \\
\hline
 error   & 0.29& 0.35  & 0.42  &0.56 & 0.58  & 0.74 &1.12    \\
\hline
\end{tabular*}
\label{tab:ForwardCrosssections}
\end{table}

\section{Summary}
\label{sec:summary}

We have reproduced the total cross section and the differential cross section
of the reaction $\gamma p \rightarrow J / \psi p$ near the production threshold
with  two-gluon exchange model encountering a parameterized gluon distribution function.
The parameterized gluon distribution function is determined by a fit to the recent GlueX data,
and it is found to be basically consistent with the global analyses of gluon distribution from other groups.
An interesting finding is that in accordance with GlueX data and within the two-gluon exchange model,
the gluon distribution does not go down quickly when $x$ approaches one.
Results inditate that
two-gluon exchange model depicts well both the differential and the total cross section of $J/\psi$ in a wide energy range,
and it can be used to predict the electroproduction cross section near the production threshold.
On EicC, the low energy EIC, the $J/\psi$ production cross section is around 10 pb based on our model,
hence EicC will be an important and interesting future machine to collect the $J/\psi$ data
and to explore the exotic hadrons in the charm sector and the nucleonic mass structure.
Assuming the integrated luminosity of EicC experiment
can reach up to 50 fb$^{-1}$ per year \cite{Chen:2018wyz,Chen:2019equ}, taking the total cross section $\sigma\simeq1.6$ pb in the energy range 4.1 GeV $< W <$ 4.6 GeV, the total number of $J/\psi$ produced on EicC then is $80000\pm14000$. Considering a detector efficiency of $20\%$ and collecting both the decay di-electrons and the decay di-muons, we are going to have about $1900\pm400$ $J/\psi$ near-threshold events detected per year. Thus, the precision of the near-threshold $J/\psi$ production experiment on EicC is promising.

\section{Acknowledgments}

X.-Y. Wang would like to acknowledge Dr. Quanjin Wang for
useful discussion and help on the experiment. We acknowledge the National Natural Science Foundation of China under Grant
No. 11705076. This work is partly supported by the HongLiu
Support Funds for Excellent Youth Talents of Lanzhou University of
Technology.

%
%

\end{document}